\documentclass{jps-cp}
\usepackage{txfonts} 
\usepackage{cite}
\usepackage{bm}
\usepackage{color}
\usepackage{color} 
\usepackage{braket,bm,comment}
\usepackage[normalem]{ulem}
\usepackage{here}

\title{Simulations of Surface X-ray Diffraction from a Monolayer $^{4}$He Film Adsorbed on Graphite}

\author{Atsuki \textsc{Kumashita}$^{1}$, Hiroo \textsc{Tajiri}$^{2\ast}$, Akira \textsc{Yamaguchi}$^{1\ast}$,  Jun \textsc{Usami}$^{2,3}$, Akihiko \textsc{Sumiyama}$^{1}$, Yu \textsc{Yamane}$^{1}$, Masaru \textsc{Suzuki}$^{4}$, Tomoki \textsc{Minoguchi}$^{5}$, Yoshiharu \textsc{Sakurai}$^{2}$,\\ and Hiroshi \textsc{Fukuyama}$^{3}$}

\inst{$^{1}$ Graduate School of Science, University of Hyogo, 3-2-1 Kouto, Kamigori-cho, Ako-gun, Hyogo 678-1297, Japan \\
$^{2}$ Japan Synchrotron Radiation Research Institute, 1-1-1 Kouto, Sayo, Hyogo 679-5198, Japan\\
$^{3}$ Cryogenic Research Center, The University of Tokyo, 2-11-16 Yayoi, Bunkyo-ku, Tokyo 113-0032, Japan\\
$^{4}$ Department of Engineering Science, University of Electro-Communications, Chofu, Tokyo 182-8585, Japan\\
$^{5}$ Institute of Physics, The University of Tokyo, 3-8-1 Komaba, Tokyo 153-8902, Japan\\
}

\email{ri21x011@stkt.u-hyogo.ac.jp, tajiri@spring8.or.jp$^{\ast}$, yamagu@sci.u-hyogo.ac.jp$^{\ast}$\\
$^{\ast}$ \text{corresponding author.}}

\recdate{}

\abst{We carried out simulations of crystal truncation rod (CTR) scatterings, i.e., one of the surface X-ray diffraction techniques with atomic resolution, from a monolayer He film adsorbed on graphite. Our simulations reveal that the 00\textit{L} rod scatterings from the He monolayer exhibit notable intensity modifications for those from a graphite surface in the ranges of approximately \textit{L} = 0.6 -- 1.7 and \textit{L} = 2.2 -- 3.5. The height of the He monolayer from the graphite surface largely affects the CTR scattering profiles, indicating that CTR scatterings have enough sensitivities to determine the surface structure of the various phases in the He layer. In particular, in the incommensurate solid phase, our preliminary experimental data show the intensity modulations that are expected from the present simulations.}

\kword{helium, low dimensional system, surface diffraction, synchrotron X-rays}

\begin{document}
\maketitle
\section{Introduction}
Atomically thin $^{3}\text{He}$ and  $^{4}\text{He}$ films physisorbed on graphite at low temperatures exhibit various two-dimensional (2D) quantum phases due to their reduced dimensionality and the large zero-point energies. Heat capacity measurement \cite{Greywall_He_HC_1993} revealed that, in the submonolayer of $^{4}\text{He}$, there exist gas, liquid, $\sqrt{3}\times\sqrt{3}$ commensurate solid (C$_{1/3}$), and incommensurate solid (IC) phases with increasing areal density. Neutron diffraction studies \cite{Carneiro_He_NS_1976,Lauter_He_ND_1980,Carneiro_He_ND_1981,Lauter_He_ND_1987,Lauter_He_NS_1991} revealed the in-plane structures of the C$_{1/3}$ and IC phases including their meltings by using super-lattice reflections. However, little is known about their structure in the transitional density regions such as the gas-liquid and C$_{1/3}$-IC transitions.

As a result of the delicate energy balance among the kinetic, He-He interaction, and He-graphite interaction, the height, $z$, of the He film from a graphite surface should be different depending on the phases. For example, in the C$_{1/3}$ phase, in which all He atoms are located on the stable points of the graphite honeycomb lattice, the height is expected to be lower than that of the IC phase. So far, however, only the height in the IC phase has been measured ($z$ = 2.85~\AA) by neutron diffraction \cite{Carneiro_He_ND_1981}.

Based on the above motivation, we started structural determination of the He films using surface X-ray diffraction (SXRD) technique, particularly crystal truncation rod (CTR) scatterings. The CTR scatterings are highly sensitive not only to structures parallel to the surface but also to those perpendicular to the surface \cite{Tajiri_SXRD_2020}. The 00\textit{L} rod scatterings offer structural information about the projected distribution of the He film with respect to the surface normal.

In this paper, we present simulation results of 00\textit{L} rod scatterings from a monolayer He film on graphite. We evaluated the expected modulations of CTR scatterings by the He layers at various heights in order to compare them with experimental data, particularly the future data associated with the phase transition. In addition to the monolayer of He, similar simulations were also performed for monolayer films of other rare gases such as Kr and Xe.


\section{Methods}
\subsection{CTR scatterings and simulation details}
\begin{figure}[tbh]
\centering
\includegraphics[width=15 cm]{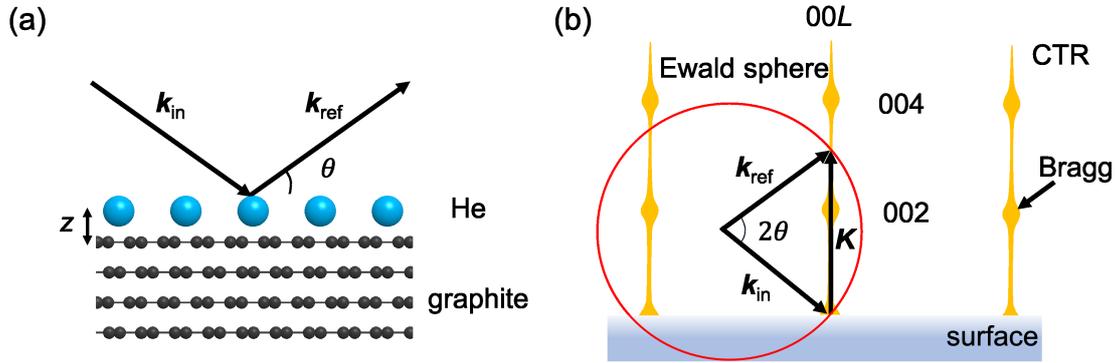}
\caption{(a) Schematics of relations between a He system and X-rays.  (b) Reflection geometry for an out of plane diffraction condition, where $\textit{\textbf{k}}_{\text{in}}$, $\textit{\textbf{k}}_{\text{ref}}$, and \textit{\textbf{K}} are the incident, reflected, and scattering wave-number vectors, respectively. Diffraction conditions are satisfied where the reciprocal lattice rods (yellow rods) intersect the Ewald sphere (a red circle). }
\label{fig:CTR_reciprocal_space}
\end{figure}

CTR scattering techniques are sensitive to surface structures at atomic resolutions. As explained in \cite{Tajiri_SXRD_2020}, when a crystal is terminated at the surface, its crystal periodicity is broken perpendicular to the surface, extending the diffraction conditions to the surface normal inbetween Bragg points. That means that X-ray scatterings are observed in the surface normal direction in addition to the Bragg reflections, which are the so-called CTR scatterings. In the Ewald construction, the diffraction conditions are satisfied where reciprocal lattice rods perpendicular to the surface intersect the Ewald sphere in the reciprocal lattice space, as shown in Fig.~\ref{fig:CTR_reciprocal_space}. Here, $\textit{\textbf{k}}_{\text{in}}$, $\textit{\textbf{k}}_{\text{ref}}$, and \textit{\textbf{K}} are the incident,  reflected, and scattering wave-number vectors, respectively. For example, the diffraction intensity, $I$, of the 00$L$ rod scattering along the reciprocal lattice unit of $L$ perpendicular to the surface is expressed by the following equation \cite{Tajiri_SXRD_2020},
  
\begin{equation}
I(L)=\left|\frac{F_{00}^{B}(L)}{1-e^{-2\pi iL}}+F_{00}^{S}(L)\right|^2,
\label{eq:CTR}
\end{equation}
where $F_{00}^{S}(L)$ and $F_{00}^{B}(L)$ are the crystal structure factors of the surface and bulk layers, respectively. The CTR intensity is expressed in terms of the interference between the X-ray scatterings from the bulk and surface. We note that the 00$L$ rod scatterings reflect only the structural information in the surface normal direction because the scattering wave number vector assumed here contains only $L$ component. The Debye-Waller (DW) factor, exp$\{-B_{j} (\text{sin}\theta/\lambda)^{2}\}$, in crystallography is involved in the crystal structure factor in the general form, $F(\bm{K})$ = $\sum_{j}f_{j}\text{exp}(2\pi i \bm{K} \cdot \bm{r}_{j})\cdot\text{exp}(-\textit{B}_{j} (\text{sin}\theta/\lambda)^{2})$, where $f_{j}$, $\bm{r}_{j}$, $\theta$, and $\lambda$ are atomic scattering factor, position vector of the $j$-th atom in a unit cell, half of the scattering angle, and wavelength of the X-rays, respectively.  The DW factor reflects  the thermal and quantum vibrations of each atom via the so-called $B$-factor, $B_{j} = 8 \pi^{2} \verb|<|u_{j}^{2}\verb|>|$, where $\verb|<|u_{j}^{2}\verb|>|$ is the mean square displacement (MSD). 


All simulations were carried out using the software for structure refinements of SXRD, SISReX \cite{Tajiri_SXRD_2020}, which also offers simulations of CTR scatterings from adsorbed layers on semi-infinite bulk layers, as expressed in Eq. (\ref{eq:CTR}). The structural parameters of both bulk and surface layers we take into account are elements on each atomic site, occupancy of the sites, positions of atoms with space group to keep crystallographic symmetry, and DW factors. Surface relaxation toward the bulk structure is also applicable,
while we did not employ it here to simplify the interpretations of the obtained results. In this paper, the surface normal components of DW factors are discussed because the characteristics of the 00$L$ rod scatterings are insensitive to in-plane structures.

     
\subsection{Synchrotron radiation experiments} 
X-ray diffraction experiments were performed at the beamline BL13XU \cite{Sakata_BL13XU_2003,Tajiri_BL13XU_2019} at SPring-8 using synchrotron X-rays of 20~keV in energy with a photon flux of 3.35×10$^{10}$~photon/s. The graphite substrate used in the experiments was a highly oriented pyrolytic graphite (HOPG) with a size of $5\times5\times0.05$~mm. Detailed experimental conditions were reported in our previous paper \cite{Yamaguchi_He_SXRD_2022}. 
\section{Simulation results}

\begin{figure}[t]
\centering
\includegraphics[width=8 cm]{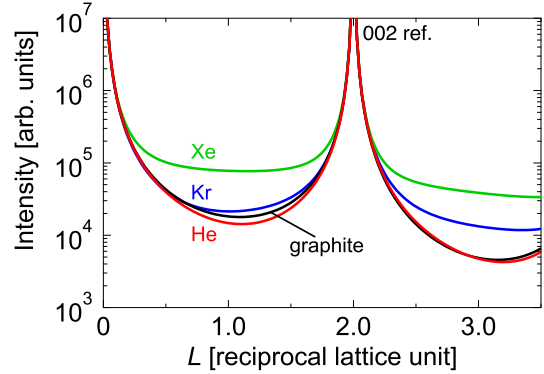}
\caption{Simulated 00\textit{L} rod scattering intensities from a $\sqrt{3}\times\sqrt{3}$ monolayer of He (red line), Kr (blue), and Xe (green) adsorbed on graphite, in addition to that from clean graphite (black). The structural parameters used in the simulations are summarized in Table \ref{tab:CTRsimu_parameter}.}
\label{fig:CTRsimu_element}
\end{figure}

As a feasibility study on whether a monolayer He film with a small atomic scattering factor is detectable using SXRD, we simulated intensities along the 00$L$ rod from monolayers of He,  Kr, and Xe on the assumption of the $\sqrt{3}\times\sqrt{3}$ commensurate phase on graphite, as shown in Fig.~\ref{fig:CTRsimu_element}. The black curve represents the calculated result for a graphite substrate without any adsorbate (clean graphite). For simplicity, slight structural relaxations of the graphite layers near the surface \cite{Yen_gr_SXRD_2004,Tajiri_Gr_SXRD_will} were omitted from our simulations, which did not affect our conclusions. Since the 00\textit{L} rod scatterings reflect the atomic distribution perpendicular to the surface ($z$-direction), the $z$ component of the $B$-factor, $B_{z}$= $8\pi^{2}\verb|<|(u_{z} )^{2}\verb|>|$, where $\verb|<|(u_{z} )^{2} \verb|>|$ is MSD along the surface normal, was considered. The $B_{z}$ value of the carbon atoms was set to $B_{z}$ = 3.19~$\AA^{2}$ \cite{Cantini_graphite_HAS_1981,Tewary_graphite_phonon_2009}. We referred to the literatures to obtain the structural parameters of the rare-gas films \cite{Carneiro_He_ND_1981,Horn_Kr_Gr_XRD_1978,Brady_Xe_Gr_XRD_1977,Cole_He_MC_1981,Shaw_Kr_LEED_1980,Cole_Xe_MC_1983}, as summarized in Table~\ref{tab:CTRsimu_parameter}. 

The simulation results are shown in Fig.~\ref{fig:CTRsimu_element}. As expected, the degrees of intensity variations from that of clean graphite roughly depend on the atomic weight of adsorbate, reflecting interferences between diffracted X-rays from the film and substrate in CTR scatterings. Nevertheless, we found relatively large intensity variations for the He film at approximately $L$ = 0.6 -- 1.7, up to a residual intensity ratio, $\Delta I/I$ of approximately minus 20~\% at most. These modifications of the intensity profile are large enough to detect with synchrotron radiation X-rays, as discussed later.

\begin{table}[b]
\centering
\caption{Structural parameters used in our simulations. $B_{z}$ of He is estimated from the density distribution reported in the literature \cite{Cole_He_MC_1981}.}
\begin{tabular}{l l l l }
\hline\hline
  & He & Kr & Xe\\ \hline
 $B_{z}\ [\AA^{2}]$& 5.29 & 4.34 \cite{Horn_Kr_Gr_XRD_1978} & 4.93 \cite{Brady_Xe_Gr_XRD_1977}\\
\textit{z} [\AA] & 2.85 \cite{Carneiro_He_ND_1981} & 3.30 \cite{Shaw_Kr_LEED_1980} & 3.35 \cite{Cole_Xe_MC_1983} \\
\hline
\end{tabular}
\label{tab:CTRsimu_parameter}
\end{table}



\begin{figure}[t]
\centering
\includegraphics[width=15 cm]{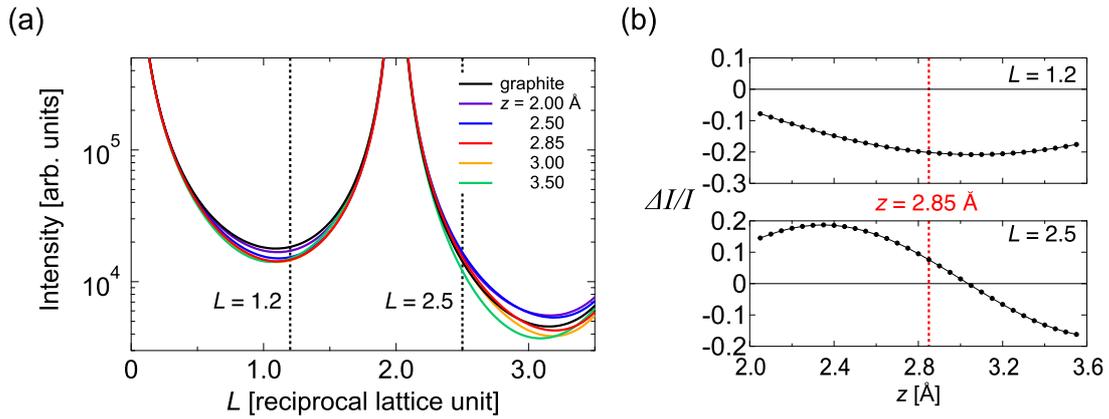}
\vskip10mm
\caption{(a) Simulated 00\textit{L} rod scattering intensities from a He monolayer in the $\sqrt{3}\times\sqrt{3}$ commensurate solid phase on graphite at various $z$ height of He. (b) Residual intensity ratio, $\Delta I/I$ of the He monolayer system from that of clean graphite as a function of the layer height $z$ at $L$ = 1.2 (upper) and $L$ = 2.5 (lower).}
\label{fig:CTRsimu_z_dens}
\end{figure}

We, then, focused on the He layer height dependence on the 00$L$ rod scatterings in the C$_{1/3}$ phase. Figure~\ref{fig:CTRsimu_z_dens} represents (a) CTR scattering profiles simulated for various $z$ and (b) the $\Delta I/I$ at $L$ = 1.2 and 2.5 as a function of $z$, respectively. As can be seen in Fig.~\ref{fig:CTRsimu_z_dens}(a), relatively large intensity variations are obtained in the ranges of approximately $L$ = 2.2 -- 3.5 as well as $L$ = 0.6 -- 1.7. Reflecting the interference effects between scatterings from the surface layer and the substrate, the intensity distribution oscillates in a complicated manner as a function of $z$. At the height of $z$ =  2.85~\AA\@ reported in the previous neutron diffraction experiment \cite{Carneiro_He_ND_1981}, a large negative value of $\Delta I/I$ was observed at $L$ = 1.2, while a relatively small but $z$-sensitive $\Delta I/I$ was observed at $L$ = 2.5, as seen in Fig.~\ref{fig:CTRsimu_z_dens}~(b). These results suggest that a synchrotron measurement at $L$ = 2.5 is more sensitive to detect the expected height deviations by the C$_{1/3}$-IC transition.

\section{Discussion}
\begin{figure}[tbh]
\centering
\includegraphics[width=8 cm]{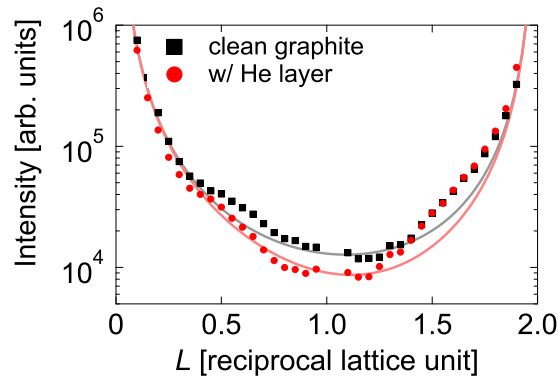}
\caption{Preliminary experimental results of 00\textit{L} rod scatterings from clean graphite (black solid squares) and He monolayer at 1.37~K with an areal density of  $\rho$ = 10.6~nm$^{-2}$ (red solid circles). Statistical errors are sufficiently smaller than the size of the plotted symbols. Simulated results are also shown by black and red lines for clean graphite and an incommensurate He on graphite at \textit{z} = 2.85~\AA, respectively.}
\label{fig:CTR_preliminary}
\end{figure}

We demonstrated that the intensity of the CTR scatterings in the C$_{1/3}$ phase (areal density of $\rho$ = 6.37~nm$^{-2}$) varies as much as minus 20~\% by the He film adsorption, which is easily observable by synchrotron X-ray diffraction. Actually, such intensity variations were confirmed in the different phases, i.e., the IC phase, by our recent observations. Figure~\ref{fig:CTR_preliminary} shows our preliminary results of the 00$L$ rod scatterings from a monolayer $^{4}$He film on graphite in the IC phase experimentally observed at $T$ = 1.37~K with an areal density of  $\rho$ = 10.6~nm$^{-2}$. We found obvious intensity modulations of the 00$L$ rod scatterings from the monolayer film (red circles) compared with those from clean graphite (black squares), especially in the range of approximately $L$ = 0.5 -- 1.5, which is attributable to He adsorption. The ratio $\Delta I/I$ = $-$29.5~\% at $L$ = 1.2 was in good agreement with our calculated value of $-$33~\% in the IC phase, in which we also assumed a He monolayer with the same areal density ($\rho$ = 10.6~nm$^{-2}$) at $z$ = 2.85~\AA. These results indicate that CTR observations of the gas-liquid transition in submonolayer He on graphite are quite feasible \cite{Yamaguchi_He_SXRD_2022}. It is noted that some discrepancies between the experimental data and our simulations might be attributed to other parameters, such as the surface relaxation of the graphite substrate. 

Regarding the possibility to observe the height difference between the C$_{1/3}$ and IC phases, we can estimate the intensity change of the CTR scattering caused by such difference as follows. According to the recent Monte Carlo calculations \cite{Gordillo_He_2009}, the energy gain in the C$_{1/3}$ phase with respect to the IC phase is approximately 3~K per atom. Since the slope $dV/dz$ of the He-graphite potential ($V$) in the first layer at $z$ = 2.85~\AA\@ is approximately 120~K/\AA \cite{Cole_He_MC_1980}, the height $z$ of the C$_{1/3}$ phase is estimated to be smaller than that of the IC phase at least by 0.025~\AA\@. This height change results in an approximately 1~\% increase of the CTR scattering intensity at $L = 2.5$, which would be detectable by a synchrotron experiment with high count statistics using high X-ray flux and enough accumulation time.


We note that, at temperatures near 1~K, the DW factors are mostly dominated by the zero-point oscillation of the atoms. This oscillation of He is much larger than that of other atoms at low temperatures, which results in a large DW factor. In fact, the large values of $B = 12.1~\AA^{2}$ \cite{Burns_bulkHe_XRD_1997} and $B = 11.8~\AA^{2}$ \cite{Blackburn_bulkHe_ND_2007} have been reported for bulk He crystals, for which the DW factor (an atomic displacement factor in the literatures) is treated as isotropic. Those for 2D He systems~\cite{Carneiro_He_ND_1981,Diallo_He_NS_2009} have also be reported. However, since anisotropy of the atomic displacement factors is crucial for arguing a confined system such as the He layer on graphite we consider, we estimated the $B_{z}$ of a $^{4}$He film as $B_{z}$ = 5.29~\AA$^{2}$, as listed in Table~\ref{tab:CTRsimu_parameter}, assuming that the MSD of the He atoms is equal to the dispersion derived by a curve fit from the reported density distribution~\cite{Cole_He_MC_1981} of the first layer. This value is much smaller than those reported for bulk He crystals~\cite{Burns_bulkHe_XRD_1997,Blackburn_bulkHe_ND_2007}, which could be attributed to the strong adsorption potential of graphite.

\section{Conclusion}
CTR scatterings from a monolayer He on graphite were discussed from a structural point of view with simulations. From our simulations, we found that the 00$L$ rod scattering intensities from a monolayer He in the C$_{1/3}$ phase vary by at most minus 20~\% compared to those from clean graphite without any adsorbate. Obvious intensity modulations of the 00$L$ rod scatterings from the monolayer He film compared to those from clean graphite were experimentally observed in good agreement with our calculations at $z$ = 2.85~\AA\@ in the IC phase, which is attributable to He adsorption. We note that the height modulations of CTR scattering intensities, even associated with gas-liquid and C$_{1/3}$-IC phase transitions, are detectable under well-defined experimental conditions. Moreover, to determine the height of the He films with high accuracy, further measurements in a wide range of reciprocal spaces are required. 
 
\section*{Acknowledgement}
This work was supported in part by JSPS KAKENHI Grant Numbers JP18H01170, JP18H03479, JP20H05621, and 22H03883. The surface X-ray diffraction measurements were performed at BL13XU and BL47XU, SPring-8 with the approval of the Japan Synchrotron Radiation Research Institute (Proposal Nos. 2021B1226, 2021B2101, and 2022A2001).


\end{document}